\documentclass[b5paper]{article}
\usepackage{times}
\usepackage{url}
\usepackage{amsmath}
\usepackage{natbib}
\usepackage{graphicx}

\begin{document}

\title{Variable Star Census in CoRoT ``Eyes''}
\author{by \\
J\'ozsef~M.~Benk\H{o}\thanks{E-mail:~benko@konkoly.hu}
  and Zolt\'an~Csubry \\
{\normalsize Konkoly Observatory,
        P.~O.~Box~67, H--1525 Budapest, Hungary}}
\maketitle


\begin{abstract}
A complete catalogue of variable stars in the possible
observing areas (``eyes'') of the CoRoT satellite is presented.
All known data sources were cross-correlated and
compiled  confirming the variability of 81 stars
formerly known as suspected variables.
By means of the TiFrAn program package a
comprehensive variability search was carried out 
on the NSVS (ROTSE-I) database for the first time in these regions. 
This search has demonstrated
the effectiveness of TiFrAn as a 
tool for finding and analysing variable stars in big databases. 
Our catalogue contains 4925 variable stars
of which 1396 stars are new discoveries. Also appended is a
list of 198 suspected variable stars.
\end{abstract}
\vskip 1mm	
\noindent
{\bf Key words:} {\it Catalogues -- Stars : variables: general
-- Surveys}

\section{Introduction}

COROT (COnvection, ROtation and planetary Transit) is a small space
mission for asteroseismology and exoplanetary search \citep{Corot}. 
In practice
it is a 30cm space telescope equipped with a 4-element mosaic CCD camera
that corresponds to two $1.4^{\circ}\times2.8^{\circ}$ 
fields-of-view: one for direct
imaging for asteroseismological purposes,
the other one is covered by an
objective prism for searching planetary transits. 
The mission can be operated in two ways: either it observes a selected 
field for $\sim$150 days (long run), or for $\sim$30 days (short run).
Annually, it is planned to have two runs of each type. The telescope 
can be pointed to two 10 degree radius circles around positions 
$\alpha=6^{\rm h}50^{\rm m}$, $\delta=0^{\circ} 0'$ (`galactic 
anticentre' or `winter' field) and
$\alpha=18^{\rm h}50^{\rm m}$, $\delta=0^{\circ} 0'$ 
(`galactic centre' or `summer' field). 
These are the CoRoT's ``eyes''.  
The satellite was launched on 27 December 2006 
and it is planned to work at least until 2009. 
Detailed  and continuously updated 
information about the project can be found    
on its  main website \url{http://smsc.cnes.fr/COROT/index.htm}

The potential targets of CoRoT asteroseismology have been extensively
studied by international collaboration and the results were uploaded to 
the GAUDI (Ground-based Asteroseismology Uniform Database Interface, see 
\citealt{GAUDI}) but the exoplanetary areas were neglected -- leastwise
from the viewpoint of variable star research.
The Berlin Exoplanet Search Telescope (BEST), a small aperture,
wide-field telescope, is dedicated to
photometric transits of exoplanets \citep{BEST} and will perform variability 
characterization of the target fields of the CoRoT mission, but
no results have been published.

CoRoT's 
exoplanetary search is aimed at obtaining light curves
from tens of thousands of stars between $\sim12$ and  $16$ mag range with
millimag accuracy and uninterrupted uniform 
sampling within 150 (or 30) days. 
These data will be unique in variable star research.   
Our main motivation was to collect present information of 
the variable star content of these areas to help in planning observations
by the satellite or related ground based ones. 

In the course of data collecting, it turned out that in the case of the
NSVS (Northern Sky Variability Survey from the ROTSE-I survey) 
database there had been no comprehensive variability search. As
is described in Section \ref{search} our search resulted in
1396 new variable stars. 

\section{Variability Search in the NSVS Database}\label{search}

The NSVS database is 
constructed from observations taken from
the first-generation Robotic Optical Transient Search Experiment
(ROTSE-I). The ROTSE-I telescope was originally designed to
find optical counterparts to gamma-ray bursters. While waiting
for a trigger from a gamma-ray burster, the telescope was set 
to patrol mode, which scanned the sky. The NSVS is based on
a year's worth of observations from 1999 to 2000.
The equipment was composed of four 20cm $f$/1.8 telephoto lenses each
equipped with $2\times2$ 
CCD mosaic cameras within $2k\times2k$ chips \citep{Keho}.
This set up gave a 14.4'' resolution corresponding to an
$8.2^{\circ} \times 8.2^{\circ}$ field-of-view for each element
of the telescope system. Based on this  
the sky was divided into 206 fields each with 
an $16.4^{\circ}\times16.4^{\circ}$ area
(see \citealt{ROTSE}). Every night the visible fields were observed in
pairs of 80 s exposures in patrol mode. In some cases,
multiple pairs of observations were taken for some fields in a single
night. The telescope did not have a filter installed so 
the sensitivity of the system is similar only to that of the 
Cousins {\it R\/} band. 

 All NSVS photometric measurements were published by \citet{Wozniak} and
are available for on-line public access from 
SkyDOT{\footnote{\url{http://skydot.lanl.gov/}}}.  
\citet{ROTSE} showed this sky patrol database to be a powerful resource 
for variable star studies. They looked over 9 sky patrol fields covering 
$\sim2000$ deg$^2$ and identified 1781 periodic variable stars. 
Although \citet{Wall} reported on 
an all sky variability census based on ROTSE-I data 
it has not been published. 
Only some specific types of variable stars were 
searched for: long period red variables \citep{Wred}, cepheids near
the equator \citep{cep}, RR Lyrae stars  \citep{Wils, Kine}, 
eclipsing binaries \citep{bin}. In view of this    
we have performed an extensive variability search in
CoRoT's ``eyes'' to complement existing catalogues.

\subsection{Preselection of Variable Star Candidates}

CoRoT's eyes are overlapped by 12 sky patrol fields 
but only in $\sim628$ deg$^2$.   
As a means of identifying whether or not 
a star in the NSVS database is a variable, 
a variability index ($I_{var}$)  was used. This index is the same as 
that outlined in \citet{ROTSE} and \citet{Kine}. 
They found this index to be well suited to the data distribution of 
ROTSE-I even though it excludes transient events such as flare stars. 
Such stars would be of interest not only from the 
viewpoint of the completeness of any variability search but 
also in their own right; however, in terms of data distribution, 
two points per night are inadequate as a means of identifying them.
 
A cut-off value had to be chosen for the
variability index in which the star could be considered a variable star
candidate. We followed \citet{ROTSE} who found
$I_{var} > 4.75 \sigma$ to be a good criterion, where $\sigma$ means 
standard deviation of the light curves. Those stars which
have less than 11 good observed points were omitted.
We extracted 
variable star candidates from the database  with these criteria
within the CoRoT eyes and obtained $\sim 82000$ candidates.

\subsection{Selection by TiFrAn}

The preselected light curves of the ROTSE-I database still 
represent a huge amount of data -- a 
quantity that cannot be managed without highly 
automated data processing algorithms. On the other hand, the low 
quality and high inhomogeneity of the data were not conductive to
automated methods. 

False signals can be caused by many effects (random or systematic errors 
of the observation, trends, sampling frequency, etc.) and it is very 
difficult to identify them automatically because of the  noisy data. 
For this reason, we applied a two-step semi-automated process to select the 
variable stars from the mass of available data.

We used TiFrAn{\footnote{The TiFrAn can be 
downloaded from \url{http://www.konkoly.hu/tifran/index.html}}}
(Time-Frequency Analyzer) program package 
\citep{tifran, tifran2} 
a scriptable data processing tool to carry out the preliminary 
time-series analysis of our data. The flexible script language of TiFrAn 
is suitable for performing complex and repetitive tasks on large 
databases.

During this automated process, we calculated the frequency spectra of 
each light curve with standard FFT (Fast Fourier Transform) method, 
and found the main frequency 
peaks in them. The {\it significance index}
 of the each peak was also calculated; for this purpose the 
following  formula was used:
\[ s = ( A_{peak} - <sp> ) / \sigma_{sp} \]
where $A_{peak}$ is the amplitude of the given peak, $<sp>$ and 
$\sigma_{sp}$ are, respectively, the average value and the standard 
deviation of the frequency spectrum. 

\begin{figure}
\centerline{\includegraphics[keepaspectratio,width=7cm]{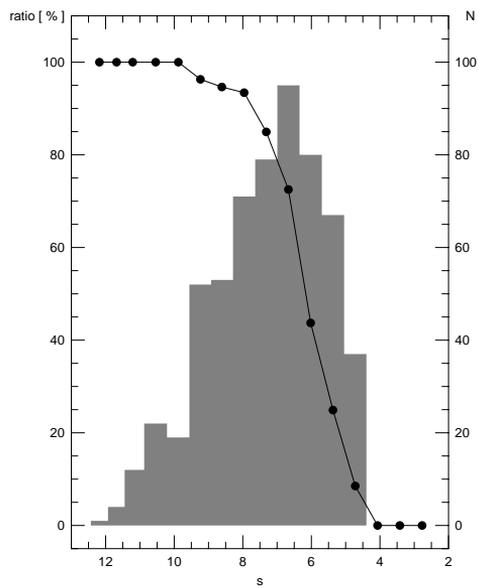}}
\caption[]{Detected number (N -- shaded histogram) and ratio 
(in per cent -- line)
of variable stars among 
preselected candidates as a function of significance index $s$.}
\label{hist}
\end{figure}

We used the significance index to narrow down the group of variable 
star candidates. A small sample  
of the full database was selected and a rough manual analysis was
performed on it to estimate the ratio of the  
variable stars among candidates with different significance indices. 
We found that this
ratio decreases rapidly with $s$ (see Fig. \ref{hist}), 
and it is less than 1\% if 
$s < 4.5$. We therefore ignored the light curves with such small $s$, 
and only the rest of the data were analysed  
(4481 stars from the winter field and 5490 from the summer field: 
about 12\% of the original data set).

\subsubsection{Long Period Variable Stars}

Before further analysis we examined the frequency distribution of the 
main peaks obtained in the Fourier spectra and 
 found that a significant number of peaks have a frequency 
value near a whole cycle per day or 
zero. As \cite{TFA} have shown, this distribution is typical
for a wide-field CCD survey and it can be 
assumed that most of these light curves belong to long period 
variable stars or otherwise they are systematic errors (trends).

To separate these two cases we tried to apply the
trend filtering algorithm (TFA) elaborated by \cite{TFA}; however,
NSVS data distribution did not allow this.
When constructing the template sets necessary for the TFA 
we need to choose a
similar amount of stars as the typical number of observing
points of a light curve.
 
\begin{figure}
\centerline{\includegraphics[keepaspectratio,width=12cm]{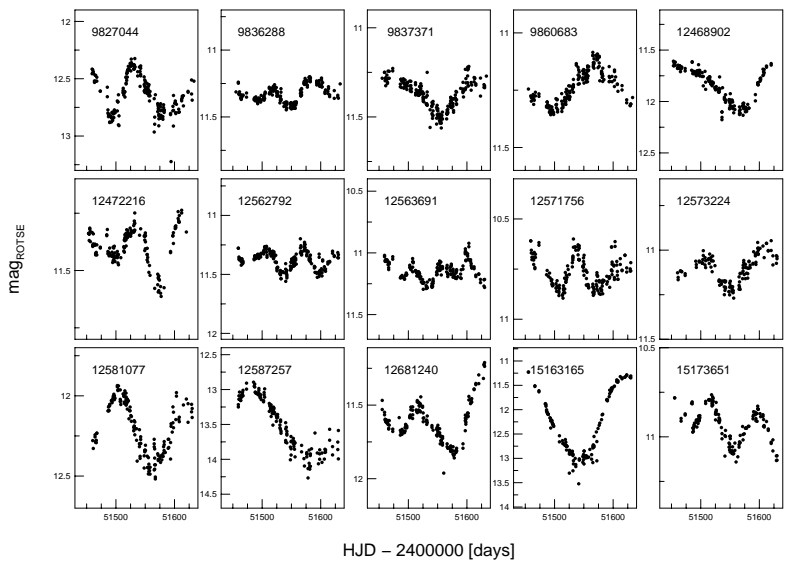}}
\caption[]{Sample from the light curves of 1338 new, long
period, semi-regular and irregular variable stars. 
The intervals between tick marks 
are the same for each panel.}
\label{long_per}
\end{figure}

Because this prevented the successful application of the method,
we  separated these light curves and manually examined them 
looking for long period variation with large amplitude. We began the 
visual examination at the light curves with the highest significance 
index.  Most of these variables can be easily verified since they 
have quite regular and large amplitude light curves. We found 2302
such variable stars of which 1338 (58~\%) were not previously known.   
A sample is shown in Fig. \ref{long_per}.

At a lower significance index the results for semi-regular and irregular 
variables became somewhat obscure. The variance of the light curve may 
originate from observational errors or trends, and in some cases 
there is no way to decide.  Therefore 198 of our selected stars from 
the low $s$ range were handled only as variable star candidates.
From these stars we prepared two separate lists, one for each
direction. 

The remaining light curves (with a frequency peak at zero or integer 
numbers, but without any sign of long-term variation) were re-processed. We 
performed a whitening on them to eliminate the sampling effects, and 
executed again the first step of our analysis. Those with significant 
peaks (with $s > 4.5$) were placed back into the data pool for further
analysis.

\subsubsection{Variable Stars with Shorter Period}

We again used the TiFrAn program package to fulfil investigations on the 
remaining data. 
The frequency and amplitude values of the main frequency peaks were 
accurately calculated using linear and 
non-linear least squares fitting methods. 
Folded light curves were also generated. 

\begin{figure}
\centerline{\includegraphics[keepaspectratio,width=12cm]{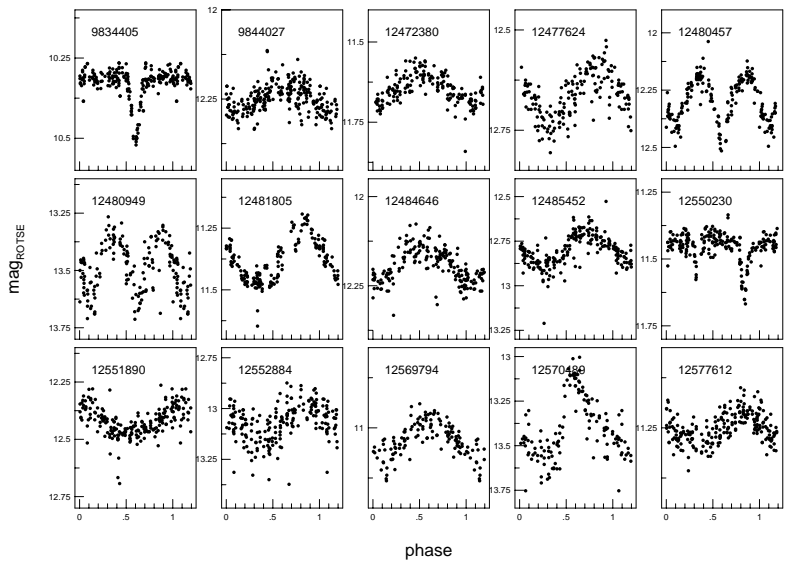}}
\caption[]{Sample phase diagrams from the set of newly discovered shorter
period variable stars.}
\label{short_per}
\end{figure}

We investigated the results of these calculations by eye in order to  
find significant evidence of the existence of periodic variation by 
visually exploring the original and folded light curves and the 
frequency spectra. With the help of this combined information, we 
successfully identified 206 shorter period pulsating variables and 
eclipsing binaries, 58 (28~\%) of these are new discoveries 
(for a sample, see Fig. \ref{short_per}).

\subsection{Variability Classification}

In accordance with the above subsections, the variable stars were divided
into three groups: the group indicated by 'L' means long period or 
not strictly periodic variables such as Miras, semi-regular or
irregular variable stars, LBVs, etc.; the group with repetitive 
light curves and shorter periods was indicated by 'S'; 
eclipsing binaries were indicated by 'E'. This last group could be
separated from
other variable stars because of their typical light curves. 
As was shown by \cite{P2002} 
pulsating variables and eclipsing binaries can be well separated
by the Fourier parameters of their light curves (see Fig. 5 in
the cited paper).

\cite{P2004} have demonstrated that the 
$\log P (J-H )$ diagram, where $P$ is the period and {\it J, H\/} are
infrared brightnesses, is a 
good tool for a first guess classification of 
pulsating stars. 
With this in mind we searched for infrared colours of
newly discovered variable stars in the 2MASS catalogue \citep{2MASS} 
and it allowed us
a more accurate classification within group 'S'.
Unfortunately, Mira and semi-regular type variables  
overlap to a considerable extent in this diagram so we 
decided not to split group 'L' into doubtful subgroups.

\subsection{Comparison with Previous Searches in NSVS}

Our variability search resulted in 2490 stars in the two 
CoRoT areas. Of this number, 1396  were not previously known. 
As mentioned in the first paragraph of  Section~\ref{search}, 
some specific searches
were carried out on this data. \cite{Wred} published 347 long
period red variable stars lying in these fields. Our pipeline missed
28 of these stars. In practically all cases this is explainable by  
the process 
having rejected all light curves with less than 11 good data points
(taking the quality flag of the ROTSE-I database into account).
On the other hand we found 1338 long period variable stars
of which a lot are red (Mira, SR).

We found the cepheid reported by \cite{cep} and an additional 6 
possible cepheids were discovered. Both RR Lyrae searches   
\citep{Wils, Kine} found no new variable stars 
in the monitored fields, but our process resulted in 10 candidates.
\cite{bin} have combined NSVS light curves with ASAS-3 \citep{P2002} data
and found 20 new eclipsing binaries. We were able to confirm 
their discoveries in 15 cases on the basis of NSVS data 
only and we found an additional 30 eclipsing binaries.

Mention is made here that because of the large scale (14.4$''/$~pixel)
and astrometric transformation problems of frames 
the synonym problem exists in the NSVS database. 
(The issue  of multiplicity is discussed by \cite{Wozniak}.)
We therefore revised our table by a shell script and
found some stars with multiple names
even in published variable list of \cite{Wred}. 
The solution to this problem was to retain the
names and positions belonging to the better light curves 
and delete the rest.

Here, we would say that our general search strategy
was successful at least with regard to the  previous studies that 
were tuned to special types of variable stars.

\section{Catalogue of Variable Stars in CoRoT ``Eyes''}

The main results of our study are four tables (two for each of the
CoRoT directions) containing all the known and presently 
discovered variable stars and new suspected ones. 
When 
preparing these tables we reviewed the literature to
find relevant sources. In addition to the NSVS, two
large databases and a number of other publications were located and 
combined. These are outlined below.

\subsection{Databases Used}

\begin{itemize}
\item
The electronic version of the Combined General Catalogue of Variable
Stars{\footnote{\url{http://www.sai.msu.ru/groups/cluster/gcvs/gcvs/}}}
(GCVS~4.2, \citealt{gcvs4.2} and further references therein) 
includes the Catalogue of Variable Stars \citep{gcvs}, 
updated and complemented with the 
Name-Lists of Variable Stars Nos.~67-78. 
This catalogue is a compilation based on a comprehensive review of the
literature and complete until 2006
but none of the databases/sources referred in this section are 
completely included.
GCVS assigns 1573 variable stars in our fields (viz. the two CoRoT eyes). 

The Combined General Catalogue also contains 
as a separate file the upgraded
New Catalogue of Suspected Variable Stars \citep{nsv} and 
its Supplement \citep{nsvsup}. This catalogue was cross-correlated
in our tables and  
58 of the suspected variable stars turned out to be real variables. 

\item
The All Sky Automated Survey (ASAS) project dedicated to the
detection and investigation of photometric variability of stars 
all over the sky. The first parts of this project
(ASAS-1 and ASAS-2) were carried out by an automated instrument 
comprising a 13.5cm telephoto lens, commercial
CCD camera and an {\it I\/} filter (for details, see \citealt{P97}).
Fifty selected $2^{\circ}\times3^{\circ}$ fields were observed between 
1997 and 2000 and the final catalogue of variable stars was published by 
\citet{P2000}. 
Two of the selected fields, S-098 and S-110, were located in CoRoT eyes 
with 207 newly discovered variable stars.

The prototype systems ASAS~1-2 were replaced by the ASAS-3 system (see
\citealt{P2002}). The ASAS-3 system 
consists of two wide-field 20cm f/2.8 instruments, 
one narrow-field 75cm f/3.3 telescope and 
one  super-wide 5cm f/4 telescope,
each equipped with an Apogee $2k\times2k$ CCD camera. 
The measurements were made in the {\it V\/} band and the
whole southern sky up to $+28^{\circ}$ declination was covered.
After a comprehensive variability search the final catalogue 
was published in five parts \citep{P2002, P2003, P2004, P2005, 
Petal05}. This is now available in electronic form as well
{\footnote{\url{http://archive.princeton.edu/~asas/}}}.
This database has provided 1192 new items for our tables.
It should be noted that even the newest electronic version 
of the ASAS-3 catalogue does not contain all ASAS~1-2 objects.
\end{itemize}

The Flagstaff Astrometric Scanning Transit Telescope (FASTT, 
\citealt{FASTT}) is
an automated 20cm meridian telescope equipped with a $2k\times2k$ CCD. 
The original aim of the FASTT survey of 16 regions arranged
along the celestial equator was to set up astrometric calibration for
the Sloan Digital Sky Survey. 
As a byproduct the survey resulted in $\sim2500$ 
new suspected variable stars (see \citealt{HS98, HS00})
with 8--15 total measurements per variable in the survey.
We cross-correlated the position of these suspected variable stars with
our catalogue and found 65 corresponding items. Twenty-one variable stars
are confirmed by our NSVS search result.  

By pursuing the goal to find 
new small amplitude variables in 
the CoRoT eyes as a secondary target of the 
asteroseismology part of the mission, photometric
observational campaigns have been started. 
The results of these campaigns are published in 
\cite{HD49434}, \cite{Por2003, Por2005}.
Our catalogue contains 61 small amplitude variables 
found by these studies. 

Some other stars were taken from smaller surveys 
such as the Misao project \citep{Mis},
Berhard's own variability survey (\citealt{Bernhard02, Bernhard03, 
Bernhard04}) or Handler's list of $\gamma$~Dor stars \citep{H99}.

Although numerous open clusters and associations are known in both
CoRoT areas 
(28 in the summer and 81 in the winter field according to the newest 
on-line version of the 
cluster catalogue of \citealt{Dias}), very few of them 
were targeted by variable star surveys.
To check in the ADS database 
among `centre' clusters, NGC~6633 was investigated 
\citep{Martin-Rodriguez, NGC6633, Hidas};  
from M11, \cite{M11} reported 39 variables.
In addition, NGC~6664, IC~4756 and Tr\"umpler~35 are clusters in
which at least one variable star member is known.
In the `anti-centre' area only NGC~2301 was investigated from
the viewpoint of  variable star research by \cite{Kim} and
\cite{Howell}, who reported 9 and $\sim2000$ variable stars, respectively,
but without positions. Thus, we could not include these stars in our
catalogue. 

It is stressed that our catalogue includes only 
optical and photometric variable stars. Neither spectroscopic variables 
without measurable photometric variability
(binaries, $\gamma$~Dor candidates, etc.)  nor non-static radio, infrared,
UV, etc. sources were added.  

\begin{table*}
\caption{ Explanation of electronic table columns}\label{t_disc}
\begin{tabular}{lll} 
\hline 
No. & Column Name & Description \\ 
\hline
1 & RA  & Right~ascension in decimal degrees ($\alpha_{2000}$) \\
2 & DEC & Declination in decimal degrees ($\delta_{2000}$)\\
3--5  & HH~MM~SS  & Right~ascension  \\
6--8  & $\pm$DD~MM~SS  & Declination  \\
9     & MAG$_{\rm ROTSE}$  & Averaged brightness from unfiltered ROTSE-I data \\
10    & PER$_{\rm ROTSE}$  & Period in days obtained from this study \\
11    & AMP$_{\rm ROTSE}$  & Fourier amplitude obtained from this study \\
12    & TYPE$_{\rm ROTSE}$  & Type of variable star \\
13    & I$_{\rm var}$  & Variability index \\
14    & ID$_{\rm ROTSE}$  & ID of star in ROTSE-I database \\
15    & ID$_{\rm ASAS}$   & ID of star in any of ASAS databases \\
16    & MAG$_{\rm ASAS}$   &  Averaged {\it V\/} or {\it I\/} brightness given by ASAS survey \\
17    & AMP$_{\rm ASAS}$   &  Total amplitude according to ASAS  \\
18    & PER$_{\rm ASAS}$   &  Period calculated by ASAS \\
19    & TYPE$_{\rm ASAS}$   &  Variable type given by ASAS  \\
20    & VARNAME   &  Variable star name from GCVS or other alternative name\\
21    & MAG$_{\rm GCVS}$   &  Maximum brightness from GCVS or average mag in \\
      &		& the referred sources in Col. 24 \\
22    & PER$_{\rm GCVS}$   &  Period given by GCVS or source Col. 24 \\
23    & TYPE$_{\rm GCVS}$   &  Type from GCVS or source Col. 24 \\
24    & REF   &  Reference number of source \\
\hline
\end{tabular}
\end{table*}

\subsection{The Catalogue}

Since our tables are rather wide no sample was inserted in this
paper. All tables of the catalogue with a detailed description 
as well as the light curves
of the new variable stars are available in electronic form via

 {\url{http://www.konkoly.hu/HAG/Science/index.html}}

Table \ref{t_disc} shows the general structure of our tables, 
viz: table\_var.centre, table\_var.anti\-centre, table\_sup.centre,
table\_sup.anti\-centre.
If one looks at Table \ref{t_disc} it might
seem that our catalogue has a number of 
redundant aspects (periods, types, magnitudes from
different sources); however, this information was given here as
a means of expanding the tables' practicableness. 
As opposed to the official GCVS values,
determination of the basic parameters of variable 
stars based upon ASAS or ROTSE data was uniform. 
Since many of the stars appeared in only one of these databases 
we decided to give the parameters from both surveys.
Furthermore, in the case of variable stars which are common in two or
three databases we get an impression of the accuracy 
of the derived parameters.   

\subsection{Distribution in the Sky}

\begin{figure*}
\centerline{\includegraphics[width=14cm]{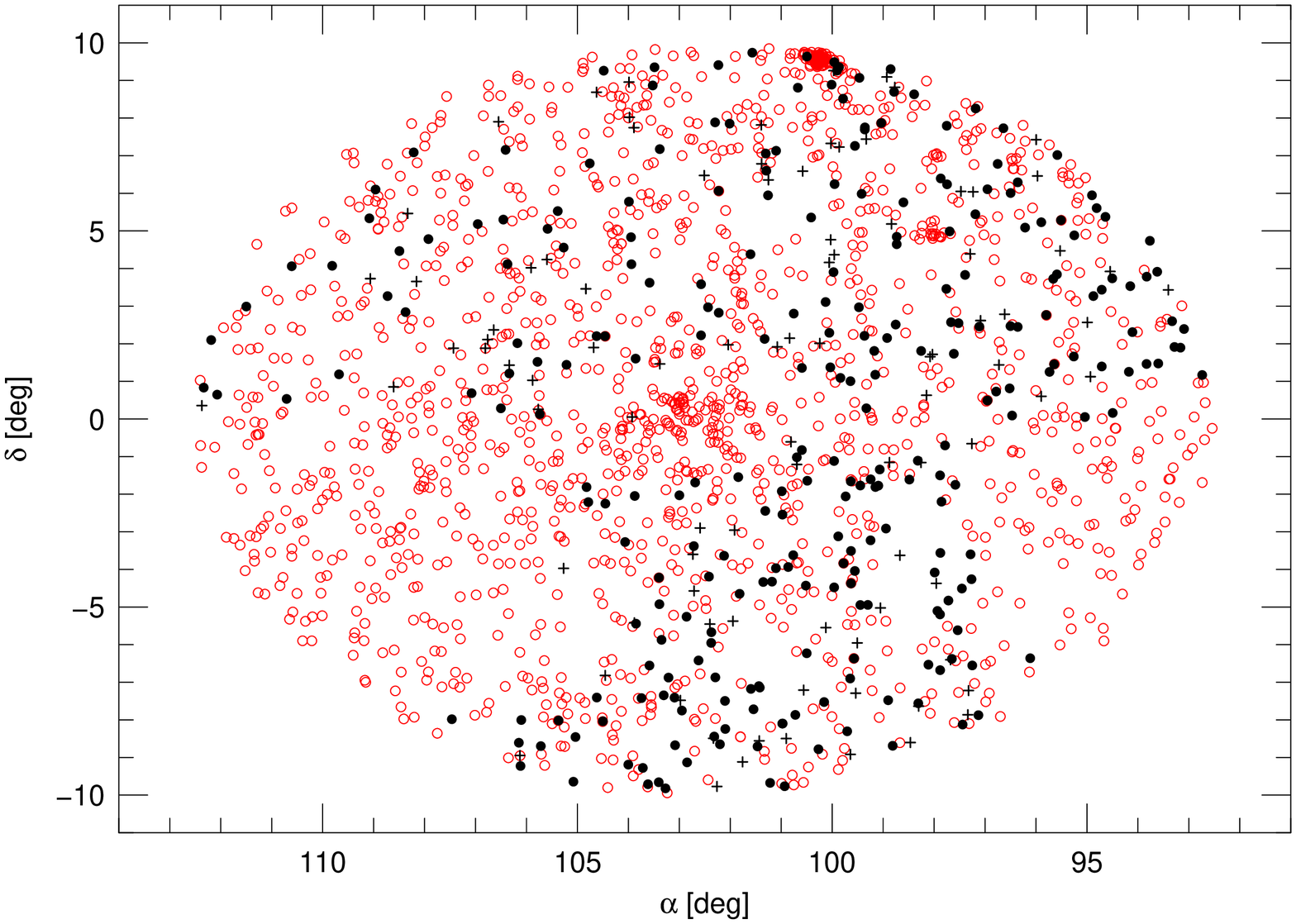}}
\caption[]{Distribution of variable stars in anticentre direction.
+: known variable stars
discovered in NSVS database; filled circles: variables found in NSVS by 
the present study; open circles: known variables from all other sources.}
\label{tel}
\end{figure*}

\begin{figure*}
\centerline{\includegraphics[width=14cm]{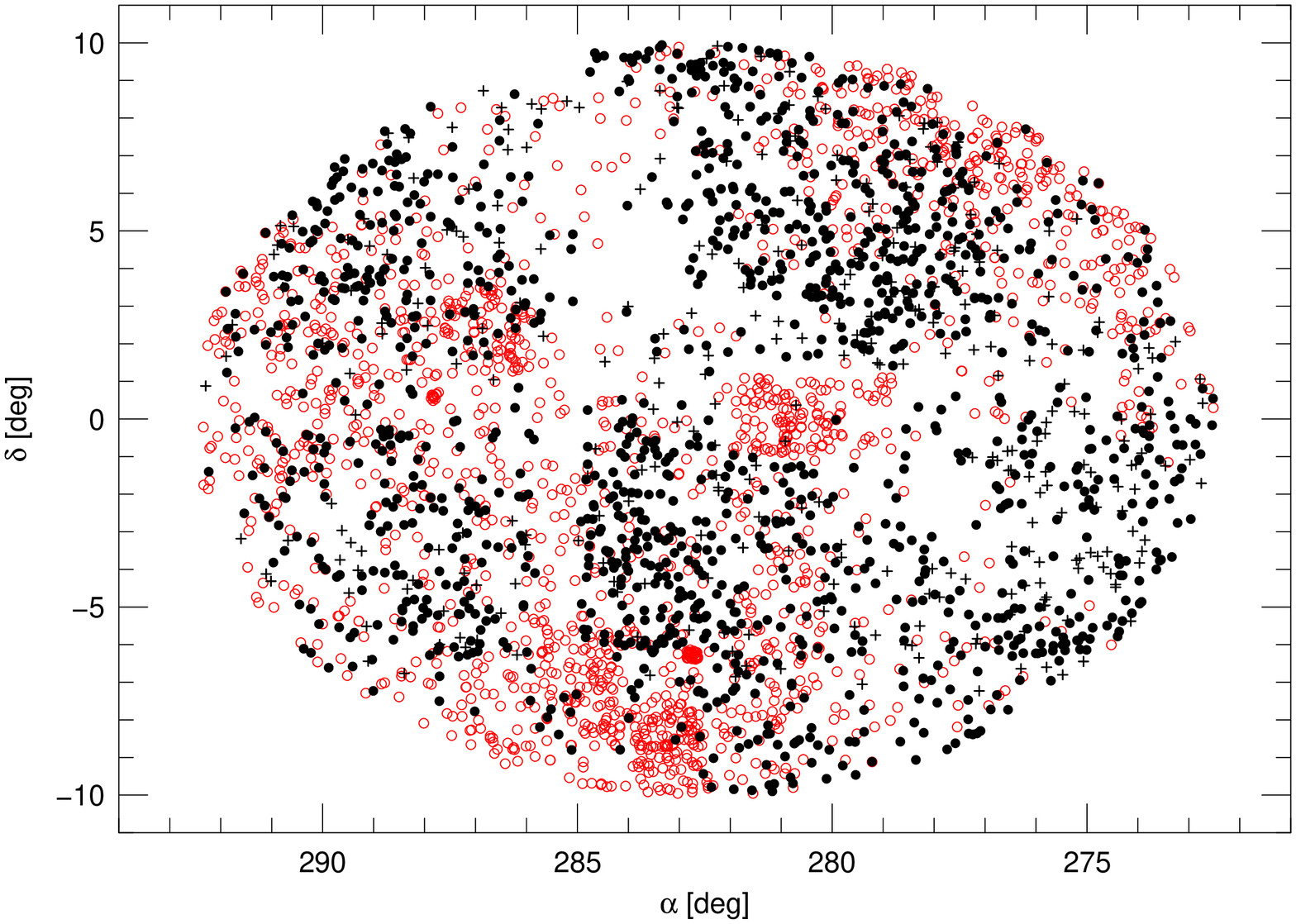}}
\caption[]{Distribution of variable stars in centre direction
(for designations, see Fig. \ref{tel}).}
\label{nyar}
\end{figure*}

If we prepare maps from the star content of our catalogue
(Figs. \ref{tel}, \ref{nyar}) it can be seen that the distribution
of these variable stars is highly non-uniform in the sky 
(in particular towards the centre).
Agglomeration of stars can be explained mainly by deeper surveys (e.g. 
ASAS~1-2, or the CoRoT team's search) but certain open clusters 
(e.g. NGC6633, M11) that were previously targeted by variable star surveys 
also appeared.
Regions with lack of stars are also due to sampling effects,
wide field surveys have generally avoided crowded regions near
the Galactic equator and/or could not present 
acceptable photometry from them (see \citealt{Wozniak, P2002}).    
Rectangular shape structures appearing in both maps are also caused 
by a similar sampling effect: they show CCD fields of surveys.
From this point of view ROTSE and ASAS surveys often 
complement each other 
(compare the distribution of different symbols in Fig. \ref{nyar}). 
Therefore, if one handles the two (NSVS and ASAS) independent data sets 
together, it generally improves only the space coverage but not the time 
one.
These factors indicate that our catalogue is far from
complete even in the moderate limit magnitude ($V \sim 14$) 
of large surveys.

\section{Summary}

We have collected such basic parameters as position, different
names, brightnesses, periods, amplitudes 
for all available variable stars in the two observing 
regions (``eyes'') of the CoRoT satellite. Beyond this compilation work
we carried out an extensive variable star search in
the NSVS database. By using the 
facilities of the TiFrAn program package, we have found 
1396 new variable stars and an additional 198 suspected variable stars.
With the help of TiFrAn, the basic 
parameters, and in the shorter period cases, the types
of these stars were also determined.

The positions of
all former variable star lists were cross-correlated 
and the light curves of corresponding items were checked for
correct identification. This work has revealed some
multiplicity in former databases and has confirmed 81 previously
suspected variable stars.

\section*{Acknowledgements}
The authors wish to thank Dr M. Papar\'o for having brought
 this topic to their  notice and for her constructive remarks and ongoing 
interest.  
This research has made use of the SIMBAD database,
operated at CDS, Strasbourg, France. It also used data from the NSVS 
and the 2MASS. The NSVS was created jointly by the LANL and the
University of Michigan. 2MASS is a joint 
project of the University of Massachusetts and the Infrared Processing and 
Analysis Center/CalTech. NSVS and 2MASS were funded by 
NASA and the NSF. The support 
provided by ESA`s PECS project is gratefully acknowledged.


\end{document}